\documentclass[conference]{IEEEtran}
\IEEEoverridecommandlockouts
\usepackage{bm}
\usepackage{cite}
\usepackage{amsmath,amssymb,amsfonts}
\usepackage{graphicx}
\usepackage{textcomp}
\usepackage{xcolor}
\usepackage{balance}
\usepackage{subcaption}
\usepackage{geometry}
\usepackage{amsthm}

\usepackage{stfloats}

\usepackage{booktabs}
\usepackage{multirow}
\begin{document}
\captionsetup[figure]{labelformat={default},labelsep=period}
\title{A Spatio-Temporal-Frequency Transformer Framework for Near-Field Target Recognition}

\author{Zongyao Zhao$^{1}$, Zhaolin Wang$^{1}$, Lincong Han$^{2}$, Jing Jin$^{2}$, Kaibin Huang$^{1}$\\
	$^1$ Department of Electrical and Computer Engineering, The University of Hong Kong, Hong Kong \\
    $^2$ Future Research Laboratory, China Mobile Research Institute, Beijing\\
	Email: \{zongyao, zhaolin.wang\}@hku.hk, \{hanlincong, jinjing\}@chinamobile.com, huangkb@hku.hk}
\maketitle
\vspace{-0.02in}
\begin{abstract}
A target recognition framework relying on near-field integrated sensing and communication (ISAC) systems is proposed. By exploiting the distance-dependent spatial signatures provided by the near-field spherical wavefront, high-accuracy sensing is realized in a bandwidth-efficient manner. A spatio--temporal--frequency (STF) transformer framework is introduced for target recognition using electromagnetic features found in the wireless channel response. In particular, a lightweight spatial encoder is employed to extract features from the antenna array for each frame and subcarrier. These features are then fused by a time-frequency transformer head with positional embeddings to model temporal dynamics and cross-subcarrier correlations. Simulation results demonstrate that strong target recognition performance can be achieved even with limited bandwidth resources.
\end{abstract}

\section{Introduction}
Integrated Sensing and Communications (ISAC) has emerged as a cornerstone technology for next-generation wireless networks\cite{Liu2022}. By sharing transceiver hardware and spectral resources for dual purposes, ISAC promises significant gains in spectral efficiency, hardware utilization, and system integration\cite{Zhao2022}. However, this convergence also introduces a core resource competition: both high-data-rate communication and high-resolution sensing inherently demand large bandwidth, while available spectrum is limited \cite{Zhao2024}.

Driven by advances in antenna technology, base stations are evolving from conventional massive MIMO to extremely large-scale antenna arrays (ELAA). As the aperture grows, many user equipment (UEs) and surrounding targets naturally lie in the radiating near field, where far-field plane-wave assumptions break down. Instead, spherical wave propagation induces range-dependent spatial signatures across the aperture and couples angle and range in the array response. This additional near-field structure enriches the spatial degrees of freedom available to both communication and sensing, and can alleviate the demands on wide bandwidth for sensing\cite{Wang2023}.

Building on this perspective, recent works have begun to investigate near-field sensing. Several studies have analyzed near-field MIMO radar localization, derived performance bounds, and designed estimators under spherical-wave propagation. In parallel, broader near-field sensing, localization, and beamforming frameworks have also been developed, together with resource allocation schemes tailored to distance-dependent channel models\cite{Wang2025,Meng2025,Zhao2025}. These efforts collectively demonstrate that near-field operation can significantly enhance sensing resolution and flexibility.

Most prior work, however, focuses on point targets. In contrast, target recognition directly from wireless sensing signals remains relatively underexplored, particularly in the near field. In practice, many objects are extended targets, whose shape, material, and orientation cause distributed scattering across the array aperture \cite{Jiang2025}. In the near field, multiple scattering centers on the same object interact through distance-dependent propagation, producing highly structured yet strongly coupled MIMO responses. Explicit parametric modeling of such signatures becomes intractable for realistic geometries and motions, limiting the applicability of model-based methods to semantic tasks like classification.

Motivated by these challenges, we study near-field sensing under strict bandwidth constraints, where only a sparse set of subcarriers is allocated for sensing. We focus on the task of near-field vehicle classification. In this regime, classical processing is inadequate: sparse subcarriers yield poor range resolution, while the channel response is a complex function of geometry, pose, motion, and material properties, making robust model-based inversion difficult.

To address this, we propose a Spatio--Temporal--Frequency (STF) Transformer framework that operates directly on the complex-valued near-field MIMO channel tensor. A lightweight spatial encoder extracts features from the antenna response of each frame and subcarrier, capturing near-field interference across the aperture. We then stack these features into a frame–subcarrier token grid and feed it to a time–frequency Transformer with positional embeddings, which fuses information over time and subcarriers. To support training and evaluation, we further construct a near-field simulation framework based on realistic 3D extended target models and full-wave electromagnetic computation. The proposed approach is grounded in electromagnetic theory and full-wave modeling, enabling physically consistent representation of extended-target scattering. Simulation results indicate that the proposed framework can maintain strong target recognition performance with a small set of subcarriers, highlighting the potential of near-field sensing under constrained bandwidth.

\emph{Notation}: In this paper, bold italic lower-case and upper-case letters denote vectors and matrices, respectively. Scalars are denoted by italic letters. The sets of real and complex numbers are denoted by $\mathbb{R}$ and $\mathbb{C}$, respectively.  $(\cdot)^T$, $(\cdot)^*$, and $(\cdot)^H$ represent transpose, complex conjugate, and Hermitian, respectively. The Euclidean norm and Frobenius norm are denoted by $|\cdot|$ and $|\cdot|_F$, respectively. The imaginary unit is $j$, satisfying $j^2 = -1$. $\mathcal{CN}(\bm{\mu}, \bm{\Sigma})$ for the complex Gaussian distribution. The differential volume element in integrals is denoted by $\mathrm{d}\bm{r}'$.The integral over a spatial domain $V$ is denoted by $\int_{V} f(\bm{r}) \mathrm{d}\bm{r}$, where $\mathrm{d}\bm{r}$ is the volume element.

The remainder of this paper is organized as follows. Sect.~\ref{sec2} introduces the signal and system model. We present the STF transformer framework in Sect.~\ref{sec3} and experimental results and discussion in Sect.~\ref{sec4}. Finally, the conclusions are drawn in Sect.~\ref{sec5}.
\begin{figure}[t]
	\centering
\includegraphics[width=0.38\textwidth]{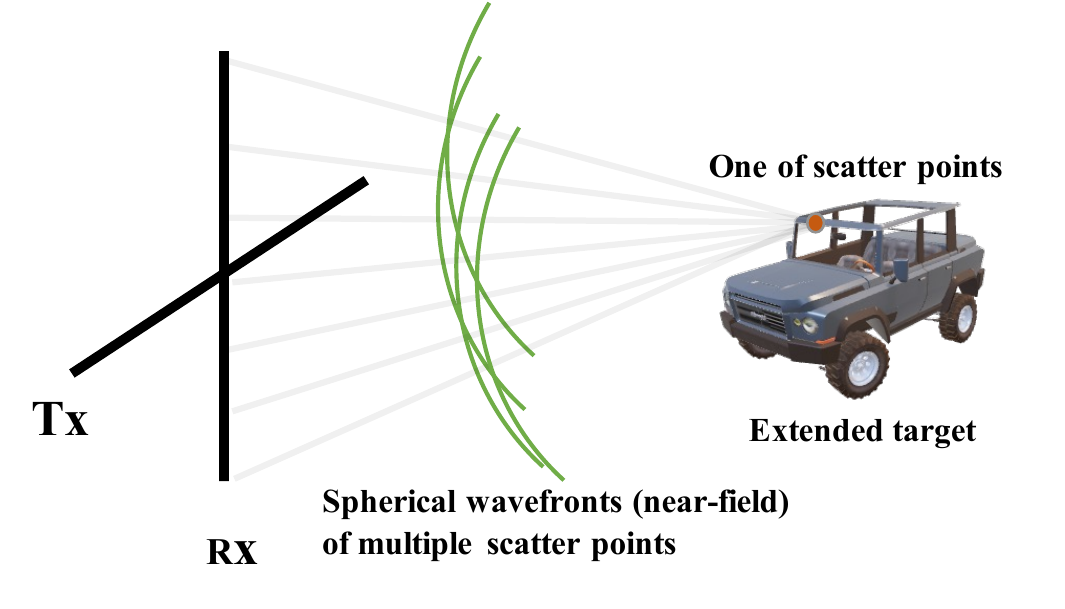}\\	\caption{Illustration of the ISAC system with an extended target in the near-field regime.}
        \label{Fig1}
        \vspace{-0.5cm}
\end{figure}
\setcounter{equation}{2}
\begin{figure*}[b]
\hrulefill
\begin{align}
    \bm E_{k}^{\mathrm{dip}}(\bm r;\bm r_{\mathrm{src}},\bm p)
    &= \frac{e^{-j k_0 R}}{4\pi \varepsilon_0}\Bigg[
       \frac{k_0^2}{R}\big(\widehat{\bm r} \times \bm p\big)\times \widehat{\bm r}
       + \Big(\frac{1}{R^3} - \frac{j k_0}{R^2}\Big)
         \big(3\,\widehat{\bm r}(\widehat{\bm r}^{\mathsf T}\bm p) - \bm p\big)
       \Bigg],
    \label{eq:dipole_incident_field_mimo}
\end{align}
\end{figure*}
\setcounter{equation}{0}
\section{System and Signal Model} \label{sec2}
We consider an ISAC base station with a cross-shaped MIMO array comprising $N_t$ transmit and $N_r$ receive antennas, as shown in Fig. \ref{Fig1}. The transmit ULA is oriented along the $y$-axis and the receive ULA along the $z$-axis, both with element spacing $d = \lambda_c/2$, providing 2D angular resolution. An extended target moves within the near-field region. Let $\bm r_t^{\mathrm{(tx)}} \in \mathbb{R}^3$ and $\bm r_r^{\mathrm{(rx)}} \in \mathbb{R}^3$ denote the  positions of the $t$-th transmit and $r$-th receive antenna elements, respectively. The transmit array lies along the $y$-axis and the receive array lies along the $z$-axis, both with electrical spacing $d=\lambda_c/2$ at the carrier wavelength $\lambda_c$. We adopt an OFDM waveform with $K$ subcarriers. On the $k$-th subcarrier, the transmitted signal vector is denoted by $\bm x_k \in \mathbb{C}^{N_t\times 1}$. The corresponding received echo at the ISAC base station is modeled as
\begin{align}\label{eq:sensing_signals}
    \bm y_k &= \bm H_k\,\bm x_k + \bm z_k,
\end{align}
where $\bm H_k \in \mathbb{C}^{N_r \times N_t}$ is the frequency-domain MIMO channel matrix (target response) on the $k$-th subcarrier, and $\bm z_k \in \mathbb{C}^{N_r\times 1}$ denotes additive noise following $\mathcal{CN}(\bm 0,\sigma_s^2\bm I_{N_r})$. In this work, our objective is to perform near-field target recognition directly from the received wireless echoes. Specifically, we assume the sensing probing vectors $\{\bm x_k\}$ are known, and the resulting MIMO target responses $\{\bm H_k\}$ (or their estimates) are used as the input features for classification.

The extended target occupies a finite region $V_s \subset \mathbb{R}^3$ with spatially varying relative permittivity $\varepsilon_r(\bm r)$ and conductivity $\sigma(\bm r)$. At angular frequency $\omega_k = 2\pi f_k$ corresponding to subcarrier $f_k$, we define the complex contrast \cite{Liu2019,Jiang2025}
\begin{align}
    \chi_k(\bm r)
    &= \varepsilon_r(\bm r) - 1
       - j\,\frac{\sigma(\bm r)}{\omega_k \varepsilon_0},
    \qquad \bm r \in V_s,
\end{align}
where $\varepsilon_0$ is the free-space permittivity and $j$ is the imaginary unit. The background medium outside $V_s$ is assumed to be homogeneous free space with permittivity $\varepsilon_0$ and permeability $\mu_0$. Each transmit element is modeled as a small electric dipole with complex dipole moment $\bm p_t \in \mathbb{C}^3$, located at $\bm r_t^{\mathrm{(tx)}}$. For a single dipole with moment $\bm p$ at position $\bm r_{\mathrm{src}}$ and an observation point $\bm r$, define
$R=\|\bm r-\bm r_{\mathrm{src}}\|$ and $\widehat{\bm r} = (\bm r - \bm r_{\mathrm{src}})/R$.
At angular frequency $\omega_k$, the electric field generated in free space is formulated as \eqref{eq:dipole_incident_field_mimo}, where $k_0 = \omega_k \sqrt{\mu_0\varepsilon_0}$ is the free-space wavenumber at subcarrier $k$ (we suppress the explicit dependence on $k$ for notational simplicity). The factor $e^{-j k_0 R}/R$ encodes spherical-wave propagation, while the $1/R^2$ and $1/R^3$ terms capture near-field reactive components.

Collecting all $N_t$ transmit elements, we form a $3\times N_t$ incident-field matrix at position $\bm r$,
\setcounter{equation}{3}
\begin{align}
    \bm A_{k,\mathrm{inc}}(\bm r)
    &= \big[
        \bm E_{k}^{\mathrm{dip}}(\bm r;\bm r_1^{\mathrm{(tx)}},\bm p_1),
        \dots,
        \bm E_{k}^{\mathrm{dip}}(\bm r;\bm r_{N_t}^{\mathrm{(tx)}},\bm p_{N_t})
       \big].
\end{align}
For a given transmit vector $\bm x_k$, the superposed incident field 
\begin{align}
    \bm E_{k,\mathrm{inc}}(\bm r)
    &= \bm A_{k,\mathrm{inc}}(\bm r)\,\bm x_k.
    \label{eq:Einc_superposition}
\end{align}

Under standard assumptions for inhomogeneous dielectric scattering, the total electric field $\bm E_k(\bm r)$ inside $V_s$ satisfies \cite{Martin1998,Vargas2022,Wang2022,Arnoldus2001,Jiang2025}
\begin{align}
    \bm E_k(\bm r)
    &= \bm E_{k,\mathrm{inc}}(\bm r)
     + k_0^2 \int_{V_s}
       \overline{\overline{\bm G}}_k(\bm r,\bm r')\,
       \chi_k(\bm r')\,\bm E_k(\bm r')\,
       \mathrm{d}\bm r',
    \label{eq:vie_total_field_mimo}
\end{align}
where $\overline{\overline{\bm G}}_k(\bm r,\bm r')$ is the free-space dyadic Green's function given by\cite{Arnoldus2001}
\begin{align}
    \overline{\overline{\bm G}}_k(\bm r,\bm r')
    &= \Big(\bm I_{3}
    + \frac{1}{k_0^2}\nabla\nabla^{\mathsf T}\Big) g_k\big(\|\bm r-\bm r'\|\big),
\end{align}
with $\bm I_{3}$ is the $3\times 3$ identity matrix and $g_k(R)$ the scalar Green's function defined as
\begin{align}
    g_k(R)
    &= \frac{e^{-j k_0 R}}{4\pi R},
    \qquad R = \|\bm r - \bm r'\|,
    \label{eq:scalar_green_mimo}
\end{align}

The scattered field at an observation point $\bm r\notin V_s$ is \cite{Vargas2022,Wang2022}
\begin{align}
    \bm E_{k,\mathrm{sca}}(\bm r)
    &= k_0^2 \int_{V_s}
       \overline{\overline{\bm G}}_k(\bm r,\bm r')\,
       \chi_k(\bm r')\,\bm E_k(\bm r')\,
       \mathrm{d}\bm r'.
    \label{eq:scattered_field_mimo}
\end{align}

Since Eq.~\eqref{eq:vie_total_field_mimo} is linear in $\bm x_k$, the total field $\bm E_k(\bm r)$ depends linearly on $\bm x_k$. Hence, for each $\bm r\in V_s$, there exists a $3\times N_t$ matrix $\bm A_k(\bm r)$ such that
\begin{align}
    \bm E_k(\bm r) = \bm A_k(\bm r)\,\bm x_k.
    \label{eq:Etot_def_A}
\end{align}
Moreover, substituting \eqref{eq:Etot_def_A} and \eqref{eq:Einc_superposition} into \eqref{eq:vie_total_field_mimo} yields an equivalent defining equation for $\bm A_k(\bm r)$:
\begin{align}
    \bm A_k(\bm r)
    &= \bm A_{k,\mathrm{inc}}(\bm r)
     + k_0^2 \int_{V_s}
       \overline{\overline{\bm G}}_k(\bm r,\bm r')\,
       \chi_k(\bm r')\,\bm A_k(\bm r')\,
       \mathrm{d}\bm r',
    \label{eq:A_VIE}
\end{align}
\begin{figure*}[t]
	\centering
\includegraphics[width=0.8\textwidth]{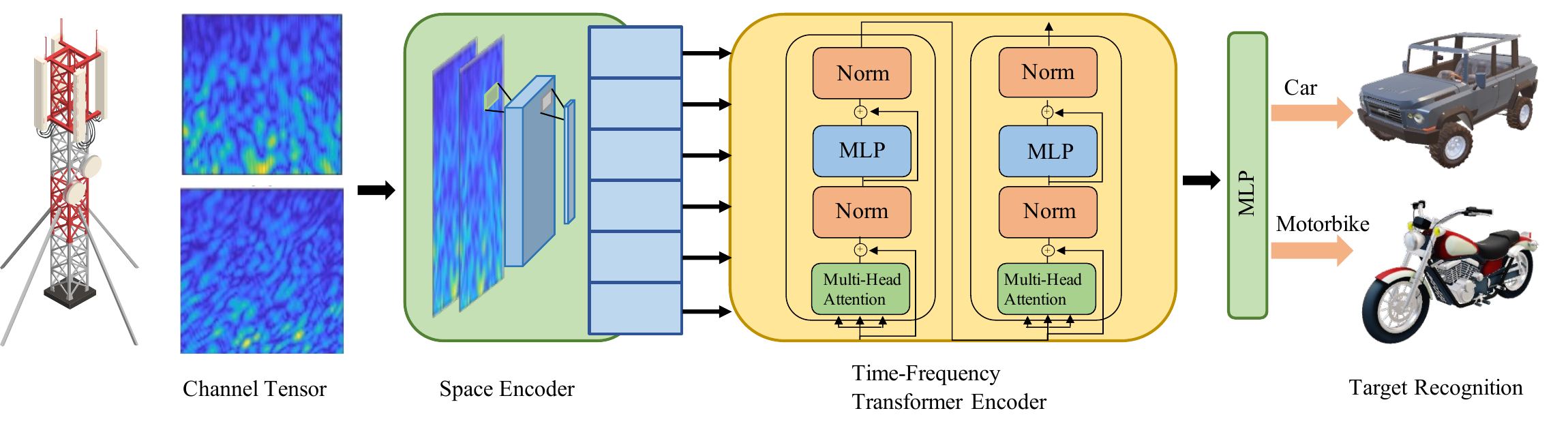}\\	\caption{Schematic of the proposed STF Transformer framework.}
        \label{Fig2}
        \vspace{-0.5cm}
\end{figure*}
Eq.~\eqref{eq:A_VIE} shows that $\bm A_k(\bm r)$ is precisely the total-field transfer matrix that maps the array excitation $\bm x_k$ to the total field at $\bm r$, and it can be obtained by solving the same VIE as \eqref{eq:vie_total_field_mimo} column-by-column.

Let $\bm q_r \in \mathbb{C}^3$ denote the receive polarization vector of the $r$-th antenna at position $\bm r_r^{\mathrm{(rx)}}$. Stacking the $N_r$ receive elements, define
\begin{align}
    \bm B_k(\bm r')
    &= \begin{bmatrix}
        \bm q_1^{H}\,\overline{\overline{\bm G}}_k\big(\bm r_1^{\mathrm{(rx)}},\bm r'\big)\\[2pt]
        \vdots\\[2pt]
        \bm q_{N_r}^{H}\,\overline{\overline{\bm G}}_k\big(\bm r_{N_r}^{\mathrm{(rx)}},\bm r'\big)
       \end{bmatrix}
       \in \mathbb{C}^{N_r\times 3}.
\end{align}
Absorbing constant antenna factors into the channel definition, the received vector on subcarrier $k$ can be written as
\begin{align}
    \bm y_k
    &= k_0^2 \int_{V_s}
       \bm B_k(\bm r')\,
       \chi_k(\bm r')\,\bm E_k(\bm r')\,
       \mathrm{d}\bm r'+\bm z_k.
\end{align}
Using \eqref{eq:Etot_def_A}, we obtain
\begin{align}
    \bm y_k
    &= k_0^2 \int_{V_s}
       \bm B_k(\bm r')\,
       \chi_k(\bm r')\,\bm A_k(\bm r')\,\bm x_k\,
       \mathrm{d}\bm r'+\bm z_k \nonumber\\
    &= \underbrace{
       k_0^2 \int_{V_s}
       \bm B_k(\bm r')\,\chi_k(\bm r')\,\bm A_k(\bm r')\,
       \mathrm{d}\bm r'
       }_{\triangleq~\bm H_k}\,\bm x_k+\bm z_k,
\end{align}
which leads to
\begin{align}
    \bm H_k
    &= k_0^2 \int_{V_s}
       \bm B_k(\bm r')\,\chi_k(\bm r')\,\bm A_k(\bm r')\,
       \mathrm{d}\bm r'.
    \label{eq:Hk_continuous}
\end{align}

Discretize $V_s$ into $N_s$ voxels with centers $\{\bm r_n\}_{n=1}^{N_s}$ and volume $\Delta V$. Then
\begin{align}
    \bm H_k
    &\approx k_0^2 \Delta V
       \sum_{n=1}^{N_s}
       \bm B_k(\bm r_n)\,
       \chi_k(\bm r_n)\,\bm A_k(\bm r_n),
    \label{eq:Hk_discrete}
\end{align}
where $\bm A_k(\bm r_n)\in\mathbb{C}^{3\times N_t}$ is the discrete total-field transfer matrix at voxel $n$ obtained from the discretized VIE, and $\bm B_k(\bm r_n)\in\mathbb{C}^{N_r\times 3}$ describes spherical-wave propagation from voxel $n$ to all receive elements.

Finally, for a sensing dwell consisting of $N_p$ slow-time frames and $K$ subcarriers, we stack the channel matrices into a fourth-order tensor
\begin{align}
    \mathcal{H} \in \mathbb{C}^{N_r \times N_t \times K \times N_p },
\end{align}
whose $(r,t,k,m)$-th entry corresponds to the channel coefficient between receive antenna $r$ and transmit antenna $t$ at subcarrier index $k$ and frame index $m$. 

With the above definitions, a sensing dwell yields the near-field channel tensor $\mathcal{H}$, which serves as the raw observation for target recognition. The recognition task is to learn a classifier that maps $\mathcal{H}$ to the target label.

\section{Spatio--Temporal--Frequency Transformer Framework} \label{sec3}
The Spatio--Temporal--Frequency Transformer framework is presented in this section and illustrated in Fig.~\ref{Fig2}.
The core objective is to learn an end-to-end mapping from the raw near-field MIMO channel tensor to target category labels, specifically designed to handle the information sparsity arising from limited sensing bandwidth.

\subsection{Data Representation and Preprocessing}

We start from the near-field channel tensor
\(\mathcal{H} \in \mathbb{C}^{N_r \times N_t \times K \times N_p}\)
obtained from the sensing process. We only use a sparse subset of subcarriers
\(\mathcal{K}_{\mathrm{sel}} \subseteq \{1,\dots,K\}\) for sensing, while the remaining subcarriers are used primarily for communication. We extract this subset to form the reduced sensing channel tensor
\begin{align}
    \widetilde{\mathcal{H}} \in \mathbb{C}^{N_r \times N_t \times K_{\mathrm{sel}} \times N_p},
\end{align}
where \(K_{\mathrm{sel}} = |\mathcal{K}_{\mathrm{sel}}|\). Each entry \([\widetilde{\mathcal{H}}]_{r,t,k,m}\) captures the complex channel response between receive antenna \(r\) and transmit antenna \(t\) at the \(m\)-th frame and the \(k\)-th sensing subcarrier.

We then decompose the complex tensor into real and imaginary parts and stack them along a new channel dimension, yielding a real-valued input tensor
\begin{align}
    \bm U \in \mathbb{R}^{2 \times N_r \times N_t \times N_p \times K_{\mathrm{sel}}}.
\end{align}
For each frame \(m\in\{1,\dots,N_p\}\) and sensing subcarrier \(k\in\{1,\dots,K_{\mathrm{sel}}\}\), we denote the corresponding antenna-domain slice by
\(\bm U_{m,k} \in \mathbb{R}^{2 \times N_r \times N_t}\).

Finally, per-sample max-magnitude normalization for each training or test sample as
\begin{align}
   \bm U \leftarrow \frac{\bm U}{M}, M \triangleq \max_{r,t,k,m} \bigl| [\widetilde{\mathcal{H}}]_{r,t,k,m} \bigr|,
\end{align}

\subsection{Network Architecture}
The STF Transformer consists of a spatial encoder for feature extraction on the MIMO array, and a time--frequency Transformer encoder for context aggregation across frames and sparse sensing subcarriers.

\subsubsection{Spatial Encoding}
The spatial encoder \(\mathcal{F}_{\mathrm{spat}}(\cdot)\) is designed to compress the high-dimensional MIMO response \(\bm U_{m,k}\) into a compact latent representation. It consists of a stack of two-dimensional convolutional layers (Conv2D) with Batch Normalization (BN) and ReLU activation. These layers scan over the \((N_r,N_t)\) aperture to capture the near-field interference fringes and wavefront curvature induced by spherical-wave propagation on the cross-shaped array.

After the convolutional stages, a global average pooling layer is applied to remove the residual spatial dimensions and produce a feature vector
\begin{align}
    \bm c_{m,k} = \mathcal{F}_{\mathrm{spat}}(\bm U_{m,k})
    \in \mathbb{R}^{D_{\mathrm{stf}}},
\end{align}
where \(D_{\mathrm{stf}}\) denotes the embedding dimension. The parameters of \(\mathcal{F}_{\mathrm{spat}}\) are shared across all \(N_p \times K_{\mathrm{sel}}\) antenna-domain slices, encouraging the network to learn spatial features that are consistent across time and frequency.

\subsubsection{Time--Frequency Transformer Encoder}

To capture the evolution of scattering features across time and the correlations across the sparse sensing subcarriers, we arrange the spatial features into a grid
\begin{align}
    \bm C \in \mathbb{R}^{N_p \times K_{\mathrm{sel}} \times D_{\mathrm{stf}}},
\end{align}
where \(\bm C[m,k,:] = \bm c_{m,k}\) stores the feature corresponding to frame \(m\) and sensing subcarrier \(k\).

We then augment each feature with learnable positional embeddings \cite{Dosovitskiy2021} that encode its temporal and spectral coordinates:
\begin{align}
    \widetilde{\bm c}_{m,k}
    = \bm c_{m,k} + \bm e^{(\mathrm{t})}_m + \bm e^{(\mathrm{f})}_k,
\end{align}
where \(\bm e^{(\mathrm{t})}_m \in \mathbb{R}^{D_{\mathrm{stf}}}\) and
\(\bm e^{(\mathrm{f})}_k \in \mathbb{R}^{D_{\mathrm{stf}}}\) are the learnable temporal and spectral embedding vectors associated with frame index \(m\) and sensing subcarrier index \(k\), respectively. These embedding vectors have the same dimension as \(\bm c_{m,k}\) and are optimized jointly with the rest of the network via backpropagation.

Stacking and flattening over the \((m,k)\) indices yields a sequence of
\(L = N_p K_{\mathrm{sel}}\) tokens
\begin{align}
    \bm T
    = \bigl[\widetilde{\bm c}_{1,1},\dots,
            \widetilde{\bm c}_{1,K_{\mathrm{sel}}},\dots,
            \widetilde{\bm c}_{N_p,K_{\mathrm{sel}}}\bigr]^{\mathsf T}
    \in \mathbb{R}^{L \times D_{\mathrm{stf}}}.
\end{align}

The token sequence \(\bm T\) is fed into a stack of \(L_{\mathrm{TF}}\) Transformer encoder layers. Each layer consists of a multi-head self-attention (MSA) block followed by a position-wise feed-forward network (FFN), with residual connections and layer normalization (LN). The MSA block captures long-range dependencies across the time-frequency grid, compensating for spectral sparsity.

Denote by \(\bm T_{\mathrm{out}} \in \mathbb{R}^{L \times D_{\mathrm{stf}}}\) the output sequence after the final Transformer layer. We then apply average pooling over the token dimension to obtain a global feature vector:
\begin{align}
    \bm g = \mathrm{AvgPool}(\bm T_{\mathrm{out}}) \in \mathbb{R}^{D_{\mathrm{stf}}},
\end{align}
which is subsequently passed through a multi-layer perceptron (MLP) classification head to produce the class logits.

\subsection{Training and Optimization}

The STF Transformer is trained in a supervised fashion using the cross-entropy loss
\begin{align}
    \mathcal{L}
    = -\frac{1}{N_B} \sum_{b=1}^{N_B} \sum_{i=1}^{N_J}
      t_{b,i} \log \hat{p}_{b,i},
\end{align}
where \(N_B\) is the batch size, \(N_J\) is the number of target classes,
\(\bm t_b \) is the one-hot target label for the \(b\)-th sample, and \(\hat{\bm p}_b\) is
the corresponding predicted class-probability vector obtained from the Softmax
output of the MLP head.

\section{Experiments and Discussion} \label{sec4}

\subsection{Dataset and Experimental Setup}
This section details the near-field simulation environment, dataset generation pipeline, and the training protocol used for evaluation.
\begin{table}[t]
    \centering
    \small
    \caption{Key simulation and sensing parameters.}
    \label{tab:exp_params}
    \begin{tabular}{ll}
        \toprule
        Parameter & Value \\
        \midrule
        \multicolumn{2}{l}{\textbf{Scenario / Channel Configuration}} \\
        Carrier frequency $f_c$ & 4.9\,GHz \\
        Number of Tx antennas $N_t$ & 64 \\
        Number of Rx antennas $N_r$ & 64 \\
        Element spacing $d$ & $\lambda_c/2$ \\
        Sensing subcarriers $K_{\mathrm{sel}}$ & $1$--$16$ \\
        Slow-time frames $N_p$ & $1$--$16$ \\
        Reference distance $R_{\mathrm{ref}}$ & 50\,m \\
        Target SNR at $R_{\mathrm{ref}}$ & 20\,dB \\
        \bottomrule
    \end{tabular}
\end{table}

\textbf{System parameters setup:}
As outlined in Section~\ref{sec2}, the simulation involves an ISAC base station and extended targets operating at a carrier frequency of $f_c=4.9$\,GHz. We employ an OFDM waveform where a subset of $K_{\mathrm{sel}}$ subcarriers is reserved for sensing, leaving the remainder available for communication. Table~\ref{tab:exp_params} summarizes the key system parameters.

\begin{table*}[t]
  \centering
  \small
  \setlength{\tabcolsep}{3pt}
  \renewcommand{\arraystretch}{1.15}
  \caption{Performance of different schemes under several selected $(N_p, K_{\mathrm{sel}})$ configurations.}
  \label{tab:stf_baselines_corner}
  \begin{tabular}{cccccccccccccccc}
    \toprule
    \multirow{2}{*}{Method} &
    \multicolumn{3}{c}{$(N_p,K_{\mathrm{sel}})=(1,1)$} &
    \multicolumn{3}{c}{$(N_p,K_{\mathrm{sel}})=(1,8)$} &
    \multicolumn{3}{c}{$(N_p,K_{\mathrm{sel}})=(4,4)$} &
    \multicolumn{3}{c}{$(N_p,K_{\mathrm{sel}})=(8,1)$} &
    \multicolumn{3}{c}{$(N_p,K_{\mathrm{sel}})=(8,8)$} \\
    \cmidrule(lr){2-4}
    \cmidrule(lr){5-7}
    \cmidrule(lr){8-10}
    \cmidrule(lr){11-13}
    \cmidrule(lr){14-16}
    & Acc & Rec & F1
    & Acc & Rec & F1
    & Acc & Rec & F1
    & Acc & Rec & F1
    & Acc & Rec & F1 \\
    \midrule
    STF  &
    $\bm{0.939}$ & $0.920$ & $\bm{0.939}$ &
    $\bm{0.984}$ & $\bm{0.976}$ & $\bm{0.984}$ &
    $\bm{0.973}$ & $\bm{0.984}$ & $\bm{0.974}$ &
    $0.961$ & $0.960$ & $0.962$ &
    $\bm{0.986}$ & $\bm{0.988}$ & $\bm{0.986}$ \\
    STF w/o TF &
    $0.904$ & $0.909$ & $0.907$ &
    $0.949$ & $0.937$ & $0.950$ &
    $0.945$ & $0.940$ & $0.946$ &
    $\bm{0.963}$ & $\bm{0.972}$ & $\bm{0.965}$ &
    $0.978$ & $0.976$ & $0.978$ \\
    FFT-CNN &
    $0.512$ & $\bm{1.000}$ & $0.677$ &
    $0.813$ & $0.770$ & $0.808$ &
    $0.880$ & $0.865$ & $0.880$ &
    $0.703$ & $0.742$ & $0.719$ &
    $0.907$ & $0.881$ & $0.906$ \\
    FFT-SVM &
    $0.512$ & $\bm{1.000}$ & $0.677$ &
    $0.728$ & $0.675$ & $0.717$ &
    $0.728$ & $0.683$ & $0.720$ &
    $0.555$ & $0.516$ & $0.543$ &
    $0.657$ & $0.595$ & $0.640$ \\
    \bottomrule
  \end{tabular}
  \vspace{-0.3cm}  
\end{table*}
\textbf{3D model setup and EM simulation:}
We generated a synthetic near-field sensing dataset using 3D point clouds of cars and motorcycles from the ShapeNet dataset \cite{ShapeNet}. Each target model was first rescaled to realistic physical size and aligned to a canonical coordinate system. The models were then embedded into the scattering volume $V_s$ and voxelized. To emulate realistic electromagnetic properties, we distinguished between metallic body panels and lossy rubber tires, assigning different piecewise-constant complex contrasts to the corresponding scatterer voxels.

Target trajectories were initialized by randomizing the starting distance $R_0 \sim \mathcal{U}(5~\text{m},50~\text{m})$ and yaw angle $\psi_0 \sim \mathcal{U}(-180^{\circ},180^{\circ})$, while ensuring that the target’s azimuth relative to the radar remained within a $\pm 60^{\circ}$ field-of-view. Each target then moved along a linear trajectory with a constant velocity $v \sim \mathcal{U}(0~\text{m/s},15~\text{m/s})$ over $N_p$ slow-time frames. Crucially, we incorporated micro-Doppler effects by leveraging ShapeNet’s semantic part annotations to isolate wheel voxels, which were rotated consistently with the vehicle’s linear velocity, thereby inducing realistic rotational micro-motion features.

Based on the processed point clouds, we solved the VIE formulated in Section~\ref{sec2} for each frame and sensing subcarrier. The resulting scattered fields were aggregated into a frequency-domain near-field channel tensor $\widetilde{\mathcal{H}} \in \mathbb{C}^{N_r \times N_t \times K_{\mathrm{sel}} \times N_p}$, labeled by the semantic category of the source 3D model.

\textbf{Network architecture:}
For input to the STF Transformer, the complex tensor $\widetilde{\mathcal{H}}$ is decomposed into real and imaginary components to form $\bm U \in \mathbb{R}^{2 \times N_r \times N_t \times N_p \times K_{\mathrm{sel}}}$, which is normalized per sample by its peak magnitude. The spatial encoder employs a lightweight Conv2D backbone (three convolutional blocks with global average pooling) to extract $D_{\mathrm{stf}}=128$-dimensional features. The subsequent time--frequency Transformer head comprises $L_{\mathrm{TF}}=2$ layers with MSA, FFN, and learnable positional embeddings, as detailed in Section~\ref{sec3}.

\textbf{Training and Baselines:}
The model is trained using cross-entropy loss and the AdamW optimizer (batch size $N_B=16$, weight decay $10^{-4}$) with cosine learning-rate decay. Model selection is based on validation accuracy. Implementation was performed in PyTorch.
\begin{figure}[t]
	\centering
\includegraphics[width=0.495\textwidth]{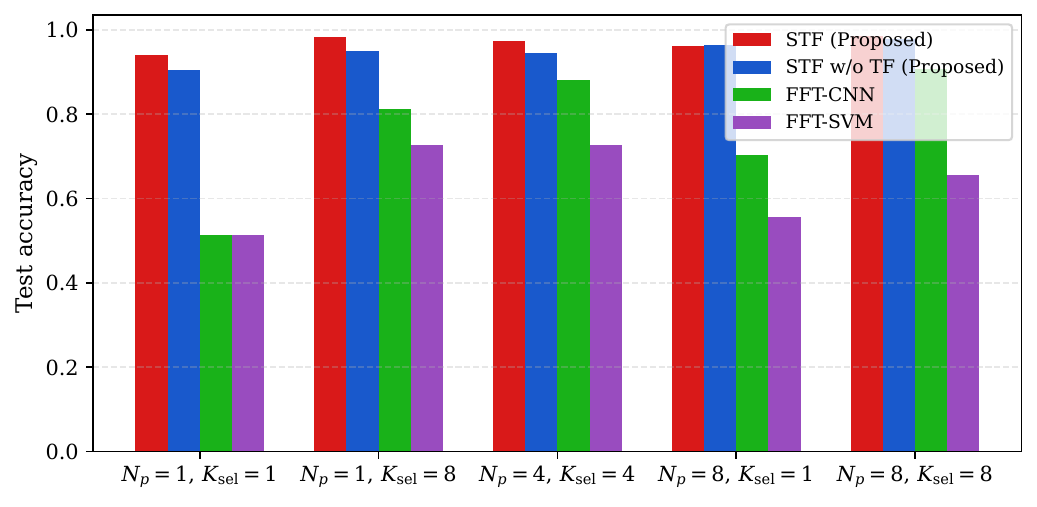}\\ \vspace{-0.3cm}	\caption{Test accuracy of STF and baselines under different sensing budgets.}
        \label{Fig3}
          \vspace{-0.5cm}   
\end{figure}
\begin{figure}[t]
	\centering
\includegraphics[width=0.4\textwidth]{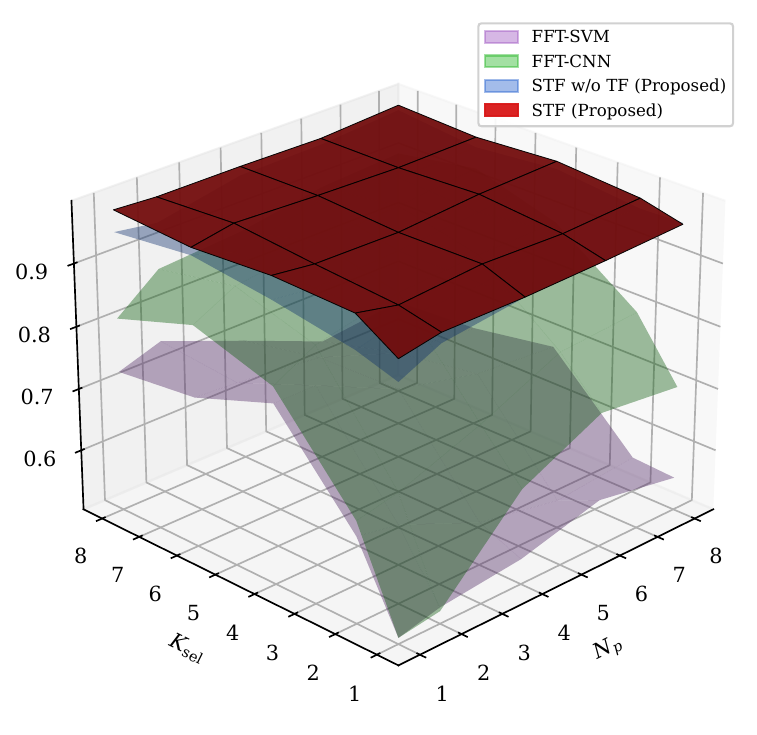}\\	  \vspace{-0.3cm}   \caption{
Test accuracy surfaces of STF and baselines}
        \label{Fig4}
          \vspace{-0.5cm}   
\end{figure}
We assess STF against three baselines. STF w/o TF replaces the time–frequency Transformer with a lightweight MLP head: the same spatial encoder is applied to $\mathbf U$ on the antenna plane, its per–$(n_p,k)$ tokens are averaged over the slow-time and subcarrier dimensions, and the resulting global descriptor is fed into a small fully connected classifier. FFT-CNN computes a physically motivated 4D FFT by applying a 2D spatial FFT over the $(N_r,N_t)$ antenna array followed by a 2D FFT over $(N_p,K_{\mathrm{sel}})$, averages over the angle dimensions to obtain a tensor, and then applies a standard 2D ConvNet to this tensor. Third, FFT-SVM uses the same 4D FFT front-end, but aggregates the tensor into a fixed-size pooled grid and trains a linear SVM on the resulting feature vector.
\subsection{Experimental Results}
Table~\ref{tab:stf_baselines_corner} summarizes the performance of the proposed STF Transformer and the baselines under several $(N_p, K_{\mathrm{sel}})$ configurations. On the test set, the STF Transformer achieves an accuracy of 98.6\%, Recall of 98.8\% and F1 of 98.6\%, outperforming baselines by a clear margin under $(N_p,K_{\mathrm{sel}})=(8,8)$. Notably, even in the extreme sparse case $(1,1)$, STF maintains 93.9\% accuracy, significantly higher than FFT-based methods. Under various configurations of $(N_p, K_{\mathrm{sel}})$, the proposed schemes (STF and STF w/o TF) consistently outperform the baseline methods, indicating that they can extract richer knowledge from the channel tensor, whereas conventional FFT-based approaches perform poorly due to the limited number of frames and small bandwidth, compounded by model mismatch.

Fig.~\ref{Fig3} and Fig.~\ref{Fig4} visualize the test accuracy of all methods as a function of the number of frames $N_p$ and sensing subcarriers $K_{\mathrm{sel}}$. The accuracy surface of STF dominates the others over most of the $(N_p,K_{\mathrm{sel}})$ grid, indicating that it can more effectively exploit additional sensing resources. The gap between proposed and the baselines becomes particularly pronounced in the low-budget region, while all methods degrade as the sensing configuration becomes extremely sparse. This plot highlights that the proposed STF architecture scales more favorably with both temporal and frequency diversity.

\begin{figure}[t]
	\centering
\includegraphics[width=0.38\textwidth]{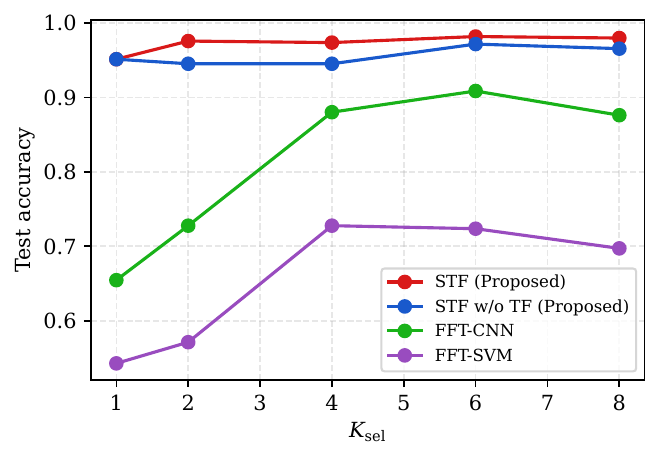}\\	
   \vspace{-0.3cm}   
\caption{
Test accuracy versus $K_{\mathrm{sel}}$ with a fixed $N_p=4$.
}
        \label{Fig5}
        \vspace{-0.3cm}   
\end{figure}

\begin{figure}[t]
	\centering
\includegraphics[width=0.38\textwidth]{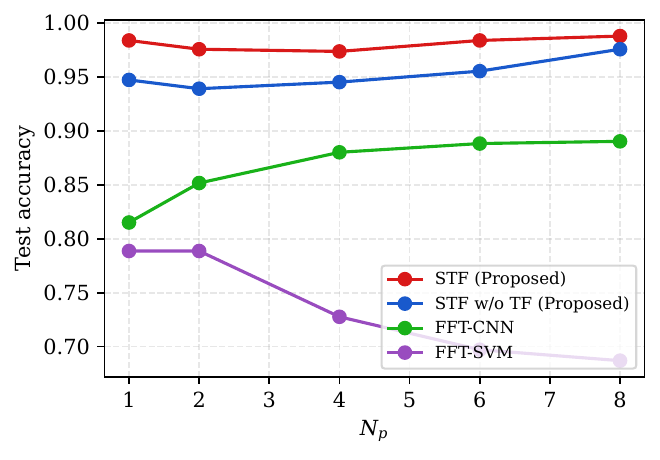}\\
  \vspace{-0.3cm}   \caption{Test accuracy versus $N_p$ with a fixed $K_{\mathrm{sel}}=4$}
        \label{Fig6}
\vspace{-0.5cm}        
\end{figure}

Fig.~\ref{Fig5} shows how the test accuracy varies with the number of sensing subcarriers $K_{\mathrm{sel}}$ when the number of frames is fixed to $N_p=4$. As $K_{\mathrm{sel}}$ generally increases, STF consistently benefits from the additional frequency diversity and maintains a clear margin over the baselines. Fig.~\ref{Fig6} examines the impact of the number of frames $N_p$ with $K_{\mathrm{sel}}$ fixed to 4. Increasing $N_p$ provides more temporal diversity and micro-Doppler information, from which STF benefits the most. 

Notably, even in the extremely sparse sensing regime with $N_p = 1$ and $K_{\mathrm{sel}} = 1$, the proposed STF-based methods still maintain strong recognition performance, clearly outperforming the baselines.

\subsection{Discussion}

The above results show that the proposed STF Transformer provides strong target classification performance in this near-field ISAC setting where only a small number of subcarriers are reserved for sensing. By first compressing the near-field MIMO response into compact spatial features, and then aggregating them across frames and sparse subcarriers via self-attention, the network effectively compensates for the lack of dense frequency sampling. From a system perspective, this suggests that near-field spatial structure may be exploited to trade sensing bandwidth for array aperture, which is attractive for ISAC designs that must prioritize communication bandwidth.

In contrast, the FFT-based baselines rely on an FFT feature that is derived from far-field. In the near-field setting with extended car and motorbike targets, this leads to noticeable model mismatch, especially as the bandwidth and number of sensing subcarriers $K_{\mathrm{sel}}$ increase. The proposed STF architecture operates directly on the raw multi-antenna time–frequency tensor without relying on far-field assumptions, scales more gracefully with the sensing budget.

\section{Conclusion}\label{sec5}
This paper investigated bandwidth-efficient near-field sensing for ISAC. We built a full-wave simulation framework based on realistic 3D car and motorbike models and generated near-field MIMO channel tensors. Based on this, we proposed a STF Transformer framework that operates directly on the complex channel data. Experiments show that STF outperforms the baselines in simulation. Notably, it still maintains strong recognition performance even with severely limited subcarriers. In practice, this suggests reserving more OFDM subcarriers for high-rate communication.

\balance


\begin{thebibliography}{00}
\bibitem{Liu2022} F. Liu, Y. Cui, C. Masouros, J. Xu, T. X. Han, Y. C. Eldar, and S. Buzzi, ``Integrated sensing and communications: Towards dual-functional wireless networks for 6G and beyond,'' {\em IEEE J. Sel. Areas Commun.}, vol. 40, no. 6, pp. 1728--1767, Jun. 2022.

\bibitem{Wang2023} Z. Wang, X. Mu, and Y. Liu, ``Near-field integrated sensing and communications,'' {\em IEEE Commun. Lett.}, vol. 27, no. 8, pp. 2048--2052, Aug. 2023.

\bibitem{Zhao2022} Z. Zhao, X. Tang, and Y. Dong, ``Cognitive waveform design for dual-functional MIMO radar-communication systems,'' in {\em Proc. IEEE Global Commun. Conf. (GLOBECOM)}, 2022, pp. 5607--5612.

\bibitem{Wang2025} Z. Wang, P. Ramezani, Y. Liu, and E. Bj\"ornson, ``Near-field localization and sensing with large-aperture arrays: From signal modeling to processing,'' {\em IEEE Signal Process. Mag.}, vol. 42, no. 1, pp. 74--87, Jan. 2025.

\bibitem{Zhao2024} Z. Zhao, {\em et al.}, ``Joint beamforming scheme for ISAC systems via robust Cramér–Rao bound optimization,'' {\em IEEE Wireless Commun. Lett.}, vol. 13, no. 3,  pp. 889--893, Jan. 2024.

\bibitem{Meng2025} C. Meng {\em et al.}, ``Near-field joint location and velocity estimation for XL-MIMO systems,'' {\em ICC 2025 - IEEE International Conference on Communications}, 2025, pp. 4523--4528.

\bibitem{Jiang2025} Y. Jiang, F. Gao, and S. Jin, ``Electromagnetic property sensing and channel reconstruction based on diffusion Schr\"odinger bridge in ISAC,'' {\em IEEE Trans. Wireless Commun.}, vol. 24, no. 8, pp. 6737--6752, Aug. 2025.

\bibitem{Zhao2025} Z. Zhao {\em et al.}, ``Joint beamforming for multi-target detection and multi-user communication in ISAC systems,'' {\em IEEE Trans. Veh. Technol.}, vol. 74, no. 9, pp. 14938--14942, Sep. 2025.


\bibitem{Liu2019} Z. Liu and Z. Nie, ``Subspace-based variational Born iterative method for solving inverse scattering problems,'' {\em IEEE Geosci. Remote Sens. Lett.}, vol. 16, no. 7, pp. 1017--1020, 2019.

\bibitem{Martin1998} O. J. F. Martin and N. B. Piller, ``Electromagnetic scattering in polarizable backgrounds,'' {\em Phys. Rev. E,} vol. 58, pp. 3909--3915, Sep 1998.

\bibitem{Vargas2022} J. O. Vargas and R. Adriano, ``Subspace-based conjugate-gradient method for solving inverse scattering problems,'' {\em IEEE Trans. Antennas Propag.}, vol. 70, no. 12, pp. 12139--12146, Dec. 2022.

\bibitem{Wang2022} F.-F. Wang and Q. H. Liu, ``A hybrid Born iterative Bayesian inversion method for electromagnetic imaging of moderate-contrast scatterers with piecewise homogeneities,'' {\em IEEE Trans. Antennas Propag.}, vol. 70, no. 10, pp. 9652--9661, 2022.

\bibitem{Arnoldus2001} H. F. Arnoldus, ``Representation of the near-field, middle-field, and far-field electromagnetic Green's functions in reciprocal space,'' {\em J. Opt. Soc. Am. B}, vol. 18, no. 4, pp. 547--555, Apr. 2001.

\bibitem{Dosovitskiy2021}
A. Dosovitskiy et al., ``An image is worth 16x16 words: Transformers for image recognition at scale,'' {\em Int. Conf. Learn. Represent.}, 2021.

\bibitem{ShapeNet} A. X. Chang {\em et al.}, ``ShapeNet: An information-rich 3D model repository,'' arXiv preprint arXiv:1512.03012, 2015.

\bibliographystyle{IEEEtran}
\end{thebibliography}
\end{document}